\documentclass[12pt]{article}

\makeatletter

\def\@authoraddress{}
\def\@title{}
\def\title#1{\gdef\@title{{\par\vskip-10pt\Large\bf
\baselineskip20pt\centering\ignorespaces\uppercase{#1}\vskip6pt}}%
\setcounter{table}{0}      \setcounter{figure}{0}
\setcounter{equation}{0}   \setcounter{section}{0}
\setcounter{subsection}{0} \setcounter{subsubsection}{0}
\setcounter{paragraph}{0}
}

\def\authors#1{\expandafter\def\expandafter\@authoraddress\expandafter
{\@authoraddress %
{\dimen0=-\prevdepth \advance\dimen0 by1.5\baselineskip
\nointerlineskip \centering
\vrule height\dimen0 width0pt\relax\ignorespaces\large\sc#1\par
}%
}%
}

\def\addresses#1{\expandafter\def\expandafter\@authoraddress\expandafter
{\@authoraddress{\nointerlineskip\vskip1pc
                 \footnotesize\it\centering\ignorespaces#1\par}}}

\def\@maketitle{%
\@title
\ifdim\prevdepth=-1000pt \prevdepth0pt\fi
\@authoraddress
}

\def\maketitle{\par
\begingroup
\let\cite\@bylinecite
\global\@topnum\z@ %
\@maketitle
\endgroup
\def\@thanks{}\def\@authoraddress{}\def\@title{}
}

\def\abstract{\par
\bgroup
\ifdim\prevdepth=-1000pt \prevdepth0pt\fi
\hsize\columnwidth
\leftskip=2em \rightskip\leftskip
\dimen0=-\prevdepth \advance\dimen0 by2pc \nointerlineskip
\noindent\vskip1.5\baselineskip\nointerlineskip\noindent\footnotesize\relax}

\newif\if@firststuff

\def\endabstract{\par
\nointerlineskip \vskip0pt
\noindent \par
\egroup
\hrule depth0pt width0pt
\global\everypar{\global\@firststufffalse}\global\@firststufftrue
}

\renewcommand\section{\@startsection {section}{1}{\z@}%
                                   {-3.5ex \@plus -1ex \@minus -.2ex}%
                                   {2.3ex \@plus.2ex}%
                                   {\normalfont\large\bfseries}}
\renewcommand\subsection{\@startsection{subsection}{2}{\z@}%
                                     {-3.25ex\@plus -1ex \@minus -.2ex}%
                                     {1.5ex \@plus .2ex}%
                                     {\normalfont\large\bfseries}}

\def\1ad{\mbox{\normalsize $^1$}}
\def\2ad{\mbox{\normalsize $^2$}}
\def\3ad{\mbox{\normalsize $^3$}}
\def\4ad{\mbox{\normalsize $^4$}}
\def\5ad{\mbox{\normalsize $^5$}}
\def\6ad{\mbox{\normalsize $^6$}}
\def\7ad{\mbox{\normalsize $^7$}}
\def\8ad{\mbox{\normalsize $^8$}}
\def\adref#1{\mbox{\normalsize $^{#1}$}}
\makeatother

\begin{document}

\def\<{\langle}
\def\>{\rangle}

\newcommand{\EQ}{\begin{equation}}
\newcommand{\EN}{\end{equation}}
\newcommand{\EQA}{\begin{eqnarray}}
\newcommand{\EQN}{\end{eqnarray}}


\title{TOWARDS COVARIANT MATRIX THEORY}


\authors{Djordje Minic\adref{1}}


\addresses{\1ad Caltech-USC Center for Theoretical Physics, Department of
Physics and Astronomy, University of Southern California, Los Angeles, CA 90089-0484 }


\maketitle


\begin{abstract}
We review an approach towards a covariant formulation of Matrix theory
based on a discretization of the 11d membrane. Higher dimensional
algebraic structures, such as the quantum triple 
Nambu bracket, naturally appear in
this approach. We also discuss a novel geometric 
understanding of the space-time uncertainty relation 
which points towards a more geometric formulation
of Matrix theory. 
\end{abstract}



\section{Introduction: Matrix vs. Membrane}

Matrix theory \cite{matrix} is conjectured to be
M-theory in the infinite momentum frame.
In this article we review an approach \cite{almy}
towards a covariant formulation of
Matrix theory based on a suitable discretization of the 11-dimensional
membrane \cite{membrane}.
Recall that Matrix theory uses $N\times N$ Hermitian matrices 
to represent the transverse coordinates (and their super-partners) of $N$
$D0$-branes  
$X^i  \quad ( i=1,2, \ldots, 9)$ \cite{joe}. The statistics of 
$D0$-branes are encoded in the 
$U(N)$ gauge symmetry ($t=X^{+}=$light-cone time) 
$X^i \rightarrow U X^i U^{-1}$.
The number $N$ of $D0$-branes is connected 
to the longitudinal momentum $P^{+}$ in the light-like direction by 
$P^{+} = {N \over R}$
where $R=g_s\ell_s$ is the compactified radius in the $X^{-}$  
direction. 

What should be the structure of a covariant Matrix theory?
We expect at least: a) a generalization of the matrix algebra and 
the emergence of higher symmetry, b) the appearance of $R$ (and perhaps $N$) as dynamical variables, 
c) a formulation of holography \cite{hol} from the ``bulk''
point of view. 

At present, the only guiding principle in trying to search for
such a structure
is that a covariant version of Matrix theory should reduce to Matrix theory if the 
light-cone gauge is chosen, 
and that it should possess 11-dimensional (super) Poincar\'{e} invariance 
in the $R\rightarrow \infty$ limit. 
We also expect the covariant
version of Matrix theory to describe many-body
interactions of 11-dimensional (super)gravitons.
Time is only globally defined in Matrix theory, so its covariant 
version should be a quantum mechanical 
theory invariant under world-line reparametrizations \cite{hadm}. Finally, the elusive transverse five-brane
should be explicitly present in such a covariant formulation.

In attempting to formulate a covariant version of Matrix theory,
it is tempting to try to discretize the 11-dimensional 
world-volume membrane theory. This is natural from the point of view
of the correspondence between 2-dimensional area preserving diffeomorphisms (APD) of 
the light-cone membrane and the Goldstone-Hoppe
large $N$ limit \cite{area} of the $U(N)$ symmetry of Matrix theory. 
In particular,
the membrane world volume is invariant under 
3-dimensional volume preserving diffeomorphisms (VPD). 
Thus one might expect that a discretized membrane 
theory should be formulated in terms of discretized VPD \cite{almy}.
We are therefore lead to the  following
pictorial correspondence 

{\large
\begin{center}
\begin{tabular}{ccc}
matrices/commutators &$\leftrightarrow$& 2D surface \\
U(N) &$\leftrightarrow$& 2D APD \\
\end{tabular}

\[\Downarrow
\]
\begin{tabular}{ccc}
triple Nambu bracket &$\leftrightarrow$& 3D volume \\
? &$\leftrightarrow$& 3D VPD \\
\end{tabular}
\end{center}}

\noindent
since the two indices of $N \times N$ matrices in Matrix theory simply 
correspond to the discretized  Fourier indices on 
the membrane, and the triple Nambu bracket generates 
3-dimensional VPD. 

In this article we first review the formal quantization problem of 3-dimensional VPD, which
underlies the covariantization approach presented in \cite{almy}, and then
discuss its physical content embodied in the space-time 
uncertainty relation \cite{stu}.

\section{Classical and Quantum Nambu Brackets}

Consider a three-dimensional space parametrized by
$\{x^i\}$.
The three-dimensional VPD on this space
are described by a differentiable map
$x^i \rightarrow y^i (x)$
such that
$\{y^1, y^2, y^3 \} =1$
where, by definition,
\EQ
\{A, B, C\} \equiv \epsilon^{ijk}\partial_i A \partial_j B \partial_k C
\label{npbracket}
\EN
is the 
Nambu triple bracket \cite{nambu}, which satisfies \cite{flato,fi,filip}
\begin{enumerate}
\item Skew-symmetry
\EQ
\{A_1, A_2,A_3\}=(-1)^{\epsilon(p)} \{ A_{p(1)}, A_{p(2)},A_{p(3)} \},
\label{skewsymmetry}
\EN
where $p(i)$ is the permutation of indices and
$\epsilon(p)$ is the parity of the permutation,
\item Derivation
\EQ
\{A_1A_2, A_3, A_4\} =A_1\{A_2, A_3,A_4\} + \{A_1, A_3,A_4\}A_2 ,
\EN
\item Fundamental Identity (FI-1) \cite{fi,filip}
\EQA
\{\{A_1, A_2, A_3\},A_4, A_5 \} +\{A_3, \{A_1, A_2,A_4\},A_5\}
\nonumber \\
+\{A_3, A_4, \{A_1, A_2, A_5\}\} =
\{A_1,A_2,\{A_3, A_4, A_5\}\}.
\EQN
\end{enumerate}

The three-dimensional VPD involve two independent functions.
Let these functions be denoted by $f$ and $g$.
The infinitesimal three-dimensional VPD generator is then given as
\EQ
D{(f, g)} \equiv \epsilon^{ijk}\partial_i f \partial_j g \partial_k
\equiv  D^k(f, g) \partial_k .
\label{3dapdgenerator}
\EN
The volume-preserving property is nothing but the identity
$\partial_k (\epsilon^{ijk}\partial_i f \partial_j
g )=0 $.
Given an arbitrary scalar function $X(x^i)$, the three-dimensional VPD
act as
\EQ
D{(f,g)}X = \{f, g, X\} .
\EN
Apart from the issue of global definition of the functions $f$ and $g$,
we can represent an arbitrary infinitesimal volume-preserving
diffeomorphism
in this form.
On the other hand, if the base three-dimensional space $\{x^i\}$
is mapped into a flat Euclidean target space of dimension $d+1$ whose
coordinates are $X^{\alpha} \, \, (\alpha =0,1,2, \ldots, d)$,
the induced infinitesimal volume element is
\EQ
d\sigma \equiv \sqrt{
\{X^{\alpha}, X^{\beta}, X^{\gamma}\}^2
}dx^1dx^2dx^3.
\EN
The volume element is of course invariant under the general
three-dimensional
diffeomorphisms.
The triple product $\{X^{\alpha}, X^{\beta}, X^{\gamma}\}$ is also
``invariant" under the VPD. More precisely, it transforms as a scalar.
Namely,
\EQ
\{Y^{\alpha}, Y^{\beta}, Y^{\gamma}\} -
\{X^{\alpha}, X^{\beta}, X^{\gamma}\} =\epsilon D(f,g)\{X^{\alpha},
X^{\beta}, X^{\gamma}\} +O(\epsilon^2)
\EN
for
$Y=X+ \epsilon D(f,g)X $.
This is due to the Fundamental Identity FI-1 which shows that the
operator $D{(f,g)}$ acts as a derivation within the
Nambu bracket. For fixed $f$ and $g$, we can define a finite transformation
by \EQ
X(t) \equiv \exp (tD(f,g)) \, \rightarrow X =
\sum_{n=0}^{\infty}
{t^n\over n!}\{f, g, \{f,g, \{ \ldots , \{f,g,\{f,g, X\}\}\ldots,\}\}\}
\EN
which satisfies the Nambu ``equation of motion" \cite{nambu}
\EQ
{d\over dt}X(t)=\{f,g, X(t)\}.
\EN
The Nambu bracket is preserved under this evolution equation.

Notice that in the case of the usual Poisson structure,
the algebra of two-dimensional area preserving diffeomorphisms is given by
\EQ
[D(f_1), D(f_2)] =D(f_3)
\EN
where
$f_3= \{f_1, f_2\}$ and
$D(f)X =\{f, X\}$.
It turns out that the three-dimensional analogue \cite{almy}
of the commutator algebra \EQ
D(A_{[1}) D(A_{2]}) =D(\{ A_1, A_2 \})
\EN
can be written using the quantum triple Nambu commutator \cite{nambu}
\EQ
[A,B,C]_N \equiv ABC-ACB+BCA-BAC+CAB-CBA
\label{nambutriple}
\EN
as follows \cite{almy}
\EQ
D(A_{[1}, A_2) D(A_{3]_N},B) = 2D(\{ A_1, A_2, A_3 \},B),
\EN
or equivalently
\EQ
D(B_{[1}, B_2) D(A_{[1},A_2) D(A_{3]_N},B_{3]_N})=
4D(\{ A_1, A_2, A_3 \},\{ B_1, B_2, B_3 \}).
\EN
Both relations are equivalent to the Fundamental Identity.
This result suggests that there is a new kind of
symmetry based on a new composition law
whose infinitesimal algebra is given by the
triple commutator (\ref{nambutriple}).
It was conjectured in \cite{almy} that this symmetry is related
to the gauge transformations that are not of the Yang-Mills type.

We turn now to the properties of the quantum Nambu bracket. What do
we mean by a quantum triple Nambu bracket? In general we want
an object $[F,G,W]$ which satisfies the properties analogous
to the classical Nambu bracket $\{f,g,w \}$ as listed above. 
(Here $f,g,w$ are
functions of three variables, and the nature of $F,G,W$ is left open
for the moment.) Thus $[F,G,W]$ is expected to
satisfy \cite{flato,fi,filip}
\begin{enumerate}
\item Skew-symmetry
\EQ
[A_1, A_2,A_3]=(-1)^{\epsilon(p)} [ A_{p(1)}, A_{p(2)},A_{p(3)} ],
\label{skewsymmetry1}
\EN
\item Derivation
\EQ
[A_1A_2, A_3, A_4] =A_1[A_2, A_3,A_4] + [A_1, A_3,A_4]A_2 ,
\EN
\item Fundamental Identity (F.I.)
\EQA
[[A_1, A_2, A_3],A_4, A_5 ] &+& [A_3, [A_1, A_2,A_4],A_5]
\nonumber \\
+[A_3, A_4, [A_1, A_2, A_5]] &=&
[A_1,A_2,[A_3, A_4, A_5]].
\EQN
\end{enumerate}

An explicit
matrix realization of the
quantum Nambu bracket, which is skew-symmetric and obeys the
Fundamental Identity was given in \cite{almy}. The construction
proceeds as follows:
Define a totally antisymmetric
triple bracket of three matrices A, B, C as
\EQ
[A,B,C] \equiv
({\rm tr} A)  [B,C] +(tr B)[C,A]+(tr C)[A,B] \label{31}.
\EN
Then ${\rm tr} [A,B,C]=0$, and if $C=1$, $[A,B,1]=N [A,B]$, where
$N$ is the rank of square matrices. 
This bracket is obviously skew-symmetric and it can be shown to 
obey the Fundamental Identity \cite{almy}.
Consider the following ``gauge transformation"
\EQ
{\delta} A
\equiv i[X,Y,A], \label{32}
\EN
where the factor $i$ is introduced for
Hermitian matrices. This transformation 
represents an obvious quantum form of the
three-dimensional volume preserving diffeomorphisms.
By the definition of
the triple bracket, the generalized gauge 
transformation takes the following
explicit form
\EQ
\delta A=i\left( [({\rm tr} X)Y-({\rm tr} Y)X, A]+({\rm tr} A)[X,Y]\right).
\label{33}
\EN
If ${\rm tr} A_i=0, i=1, ... n$, then
${\rm tr} (A_1A_2\dots A_n)$
is gauge invariant.
The gauge transformation (\ref{33})
indicates that a bosonic
Hermitian matrix $A$  can be transformed into a form proportional to
the unit $N \times N$ matrix as long as ${\rm tr} A\ne 0$. In other
words, since the gauge transformation is traceless,
one can show that 
$A\rightarrow {1 \over N}({\rm tr} A) {{\bf 1}_{N \times N}}.$

The triple quantum Nambu bracket can 
also be represented in terms of
three-index objects or cubic matrices \cite{almy}.
These objects might be relevant for the description
of the five-brane degrees of freedom.

The explicit example of the quantum Nambu bracket 
presented above does not satisfy the derivation property. 
Hence, this form of the quantum 
Nambu bracket does not completely parallel
the form
of the classical Nambu bracket.
For example, the classical triple Nambu bracket of 
three functions $f, g, h$ of three 
variables $\tau,\sigma_1, \sigma_2$ can be obviously rewritten as
$
\{ f, g, w \} = \dot{f} \{g,w\} + \dot{g} \{w,f\} + \dot{w} \{f,g\}$
where $\dot{f}= \partial_{\tau} f$ and
$\{f,g\} = \partial_{\sigma_1} f \partial_{\sigma_2} g
- \partial_{\sigma_2} f \partial_{\sigma_1} g$.
If we try to extrapolate our previous definition of the 
quantum Nambu bracket in terms of $\tau$-dependent square matrices to
$
[ F, G, W ] = \dot{F} [G, W] + \dot{G} [W, F] + \dot{W} [F, G]
$
where $\dot{F}= \partial_{\tau} F$,
one can show that such $[F, G, W]$ does not satisfy the Fundamental Identity. 

Finally, we note that the formal quantization problem of Nambu brackets was
solved using the Zariski deformation quantization, based on factorization of
polynomials in several real variables \cite{flato}. Unfortunately,
a matrix relization of this quantization procedure is not known.
Thus a discretized version of the 11-dimensional membrane
based on the triple Nambu bracket is still
unavailable.

To make progress, one should perhaps consider (in
the $N \rightarrow \infty$ limit), instead of
Hermitian $N \times N$ matrices, selfadjoint operators on a Hilbert space, as is customary in Connes' quantized
calculus, and apply this machinery to the Polyakov action
for the 11-dimensional membrane following \cite{connes}.

\section{Space-Time Uncertainty Relation and Geometry of M-theory}

In a variety of contexts in perturbative and non-perturbative string theory
the following space-time uncertainty
relation \cite{stu} appears 
\EQ
\delta T \delta X \sim \alpha' . \label{str}
\EN
Here $\delta T$ and $\delta X$ measure the appropriate longitudinal 
and transverse space-time distances. Although eq. (\ref{str}) 
is nothing but the usual
energy-time uncertainty relation applied to string theory, in which
$\delta E \sim {\delta X}/{l_s^2}$, this relation appears in different
ways in different aspects of string theory.
In perturbative string theory the space-time 
uncertainty relation stems from conformal symmetry \cite{stu}.
In non-perturbative string/M theory, the space-time uncertainty 
relation captures the essential features of the physics of D-branes 
as well as the properties of holography \cite{hol} and the $UV/IR$ relation
\cite{uvir}. 
Eq. (\ref{str}) is true in Matrix theory if
$X^{a} \rightarrow \lambda X^{a}, t \rightarrow {\lambda}^{-1} t, \label{xt}$
provided the longitudinal distance
is identified with the global time of Matrix theory
and provided the string coupling 
constant is simultaneously rescaled
$g_s \rightarrow {\lambda}^{3} g_s $.
The space-time uncertainty relation hence leads readily to the well known
characteristic space-time scales in M-theory \cite{dkps}.

The string theory space-time uncertainty relation can also be
understood as a limit of a space-time uncertainty relation in M-theory 
\cite{stu}. 
In Matrix theory eq. (\ref{str}) can be rewritten as
$\delta T \delta X_{T} \sim l_{p}^{3}/R $
where $\delta X_{T}$ and $\delta T$ 
respectively measure transverse spatial and time directions. 
Note that the uncertainty for the longitudinal direction in physical
processes that involve individual $D0$-branes is 
$\delta X_{L} \sim R$. Thus
\EQ
\delta T \delta X_{T} \delta X_{L} \sim l_{p}^{3} . \label{mst1}
\EN
This is the space-time uncertainty relation in M-theory. Note that this
``triple'' relation naturally invokes the form of the triple Nambu bracket,
discussed in section 2, thus providing another physical motivation for the
covariantization approach of \cite{almy}.

Here we want to comment on the geometrical 
structure underlying the space-time uncertainty relation
which points towards a more geometric formulation
of Matrix theory. 

Recall the remarkable fact 
that the projective Hilbert space ${\cal P}$, defined as the set of rays
of the Hilbert space ${\cal H}$, for any quantum mechanical system is
a K\"{a}hler manifold \cite{gqm}. States of a quantum mechanical system 
are represented as points of this manifold.
The Schr\"{o}dinger dynamical evolution
is represented by the symplectic flow generated by a Hamiltonian function.
The distance along a given curve in the projective Hilbert space is
given by the Fubini-Study metric.
In particular, the infinitesimal distance defined by the line element 
of the Fubini-Study metric is proportional to the energy-time uncertainty
\cite{gqm}!

Let us apply this beautiful geometric formulation of quantum mechanics
to a very particular quantum mechanics of gravity - Matrix theory.
By reinterpreting the
space-time uncertainty as the energy-time uncertainty, we have the claim: 
{ \it the space-time
uncertainty in Matrix theory measures the infinitesimal distance along
a given curve in the projective Hilbert space of Matrix theory}.

Recall that the Hilbert space of Matrix theory can be represented
in terms of block diagonal $N \times N$ Hermitian matrices (at large $N$),
according to \cite{matrix}. This fact offers a tantalizing possibility
that the kinematical set-up as described by the geometric formulation
of quantum mechanics \cite{gqm} can be naturally generalized in the case of
quantum mechanics of gravity: the points of
the symplectic manifold of that quantum mechanics can be turned into matrices
and the Hamiltonian formulation of the Schr\"{o}dinger evolution
can be generalized to incorporate a non-Abelian structure!

Obviously, a geometrical reinterpretation of Matrix theory is
called for in view of the geometric meaning of the space-time
uncertainty relation.
One possible geometric interpretation might be provided by an M-theoretic generalization
of the geometric formulation of perturbative string theory
as presented some years ago in \cite{loop}.
One particularly striking 
result of that work was that the equations of motion for the massless
fields of the closed string theory were obtained from the existence
of a reparametrization invariant symplectic structure on the loop space.
This was a geometric reinterpretation of the usual requirement of 
conformal invariance in perturbative string theory.

It is natural to conjecture that {\it the equations of motion of the
11-dimensional supergravity follow from 
the existence of a $U(\infty)$ invariant 
simplectic structure of the appropriate
K\"{a}hler manifold associated with the projective Hilbert 
space of Matrix theory}. This statement would be the Matrix theory 
analogue of the requirement of conformal invariance in
perturbative string theory.

We hope to address these fascinating issues in detail elsewhere.

\noindent{\bf Acknowledgements.} It is my pleasure to thank 
V. Balasubramanian, T. Banks, I. Bars, R. Corrado, 
M. Douglas, J. Gomis, R. Gopakumar, 
T. H\"{u}bsch, R. Leigh, N. Nekrasov, S. Shenker,
L. Smolin and E. Witten  
and especially my collaborators H. Awata, M. Li and T. Yoneya 
for many interesting
discussions.
This work is supported in part by DOE grant
DE-FG03-84ER40168.

\end{document}